\documentclass[twocolumn]{webofc}
\usepackage[varg]{txfonts} 
\usepackage{graphicx}
\usepackage{amsmath}
\usepackage{amssymb}

\begin{document}

\title {New developments in aerosol measurements using stellar photometry}

\author{\firstname{Jan} \lastname{Ebr}\inst{1}\fnsep\thanks{\email{ebr@fzu.cz}} \and
\firstname{Jakub} \lastname{Jury\v{s}ek}\inst{1} \and
\firstname{Michael} \lastname{Prouza}\inst{1} \and
\firstname{Ji\v{r}\'{i}} \lastname{ Bla\v{z}ek}\inst{1} \and
\firstname{Petr} \lastname{ Tr\'{a}vn\'{i}\v{c}ek}\inst{1} \and
\firstname{Du\v{s}an} \lastname{Mand\'{a}t}\inst{1} \and
\firstname{Miroslav} \lastname{Pech}\inst{1}\ 
for the Pierre Auger Collaboration\thanks{Full author list: http://www.auger.org/archive/authors\_2018\_09.html} and the CTA Consortium\thanks{http://www.cta-observatory.org/}
\and 
\firstname{Petr} \lastname{Jane\v{c}ek}\inst{1}\  for the CTA Consortium\footnotemark[3]
\and
\firstname{Sergey} \lastname{Karpov}\inst{1} \and
\firstname{Ronan} \lastname{Cunniffe}\inst{1} \and
\firstname{Martin} \lastname{Ma\v{s}ek}\inst{1} \and
\firstname{Ji\v{r}\'{i}} \lastname{ Eli\'{a}\v{s}ek}\inst{1,4} \and
\firstname{Martin} \lastname{Jel\'{i}nek}\inst{2} \and
\firstname{Ivana} \lastname{Ebrov\'{a}}\inst{1,3}
}

\institute{Institute of Physics of the Czech Academy of Sciences, Prague 
\and
Astronomical Institute of the Czech Academy of Sciences, Ond\v{r}ejov
\and
Nicolaus Copernicus Astronomical Center, Polish Academy of Sciences, Warsaw
\and Faculty of Mathematics and Physics, Charles University, Prague
}

\abstract{
The idea of using stellar photometry for atmospheric monitoring for optical experiments in high-energy astrophysics is seemingly straightforward, but reaching high precision of the order of 0.01 in the determination of the vertical aerosol optical depth (VAOD) has proven difficult. Wide-field photometry over a large span of altitudes allows a fast determination of VAOD independently of the absolute calibration of the system, while providing this calibration as a useful by-product. Using several years of data taken by the FRAM (F/(Ph)otometric Robotic Atmospheric Monitor) telescope at the Pierre Auger Observatory in Argentina and about a year of data taken by a similar instrument deployed at the planned future Southern site of the Cherenkov Telescope Array in Chile, we have developed methods to improve the precision of this measurement technique towards and possibly beyond the 0.01 mark. Detailed laboratory measurements of the response of the whole system to both the spectrum and intensity of incoming light have proven indispensable in this analysis as the usual assumption of linearity of the CCD detectors is not valid anymore for the conditions of the observations.}

\maketitle

\section{Introduction}

In the previous AtmoHEAD conference, we have provided a detailed description \cite{VAOD} of the method to measure the vertical aerosol optical depth (VAOD) by comparing the apparent brightness $m_\mathrm{inst}$ (in magnitudes) of a large amount of stars observed on a series of images from a wide-field telescope with the expected value $m_\mathrm{cat}$ extracted from a catalog. These images are usually taken to cover a wide range of airmasses $A$ in a vertical or slanted ``scan''. The overall atmospheric extinction is extracted from a fit of the data with a model that includes both atmospheric and instrumental effects and can be written, for each star, as:
\begin{equation}
\begin{split}
m_\mathrm{inst}=Mm_\mathrm{cat}+Z_i+k_iA+c_1(B-V)(c_2(B-V)+1)+ \\
R_1r(R_2r+1)+k_CA(B-V)+k_{A2}A^2\label{model}
\end{split}
\end{equation}
where $B-V$ is the color index of the star, $r$ its distance from the center of the chip, $M,c_1,c_2,R_1,R_2,k_C,k_{A2}$ are global parameters that depend on the experimental setup, $Z_i$ is the calibration constant (zeropoint) for each scan and $k_i$ is the extinction coefficient from which the VAOD can be deduced after subtracting the effects of the molecular atmosphere. 

Currently, there are two such wide-field telescopes (FRAMs -- F/(Ph)otometric Robotic Atmospheric Monitors) in operation: at the Pierre Auger Observatory \cite{AUGER} near Malarg\"{u}e, Argentina and at the future planned Southern site of the Cherenkov Telescope Array (CTA) \cite{CTA} near Cerro Paranal, Chile (see \cite{PJ} for details). As the atmospheric conditions on these sites, as well as the hardware setups, are quite different, comparing the results from the two FRAMs brings valuable insights for the further development of the method. The molecular contribution and the $k_{A2}$ parameter must be obtained from the spectral response of the system which has been, for every combination of hardware used at any site, measured in the optical laboratory in Olomouc (the $k_C$ parameter can be either calculated or fitted from the data). Note that all of the measurements presented here are taken in the photometric B filter at an effective wavelength for VAOD determination between 430 and 440 nm depending on the type of aerosols and the specific hardware setup.

\begin{figure*}
\begin{centering}
\includegraphics[width=\textwidth]{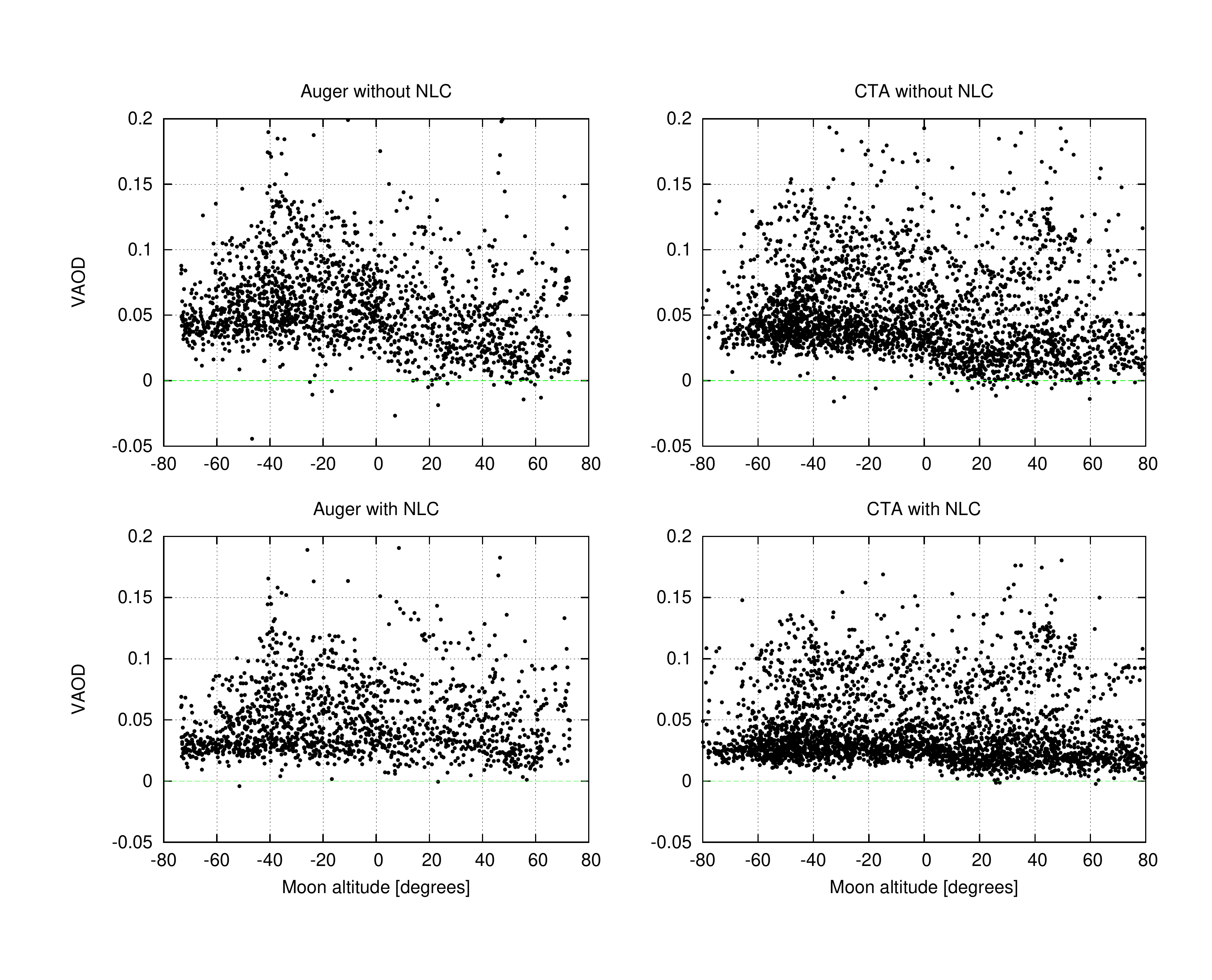}
\caption{The dependence of the measured VAOD on the Moon altitude above the horizon for the Auger (left) and CTA (right) FRAMs. The top row shows the results before the non-linearity correction, the bottom row shows results with the non-linearity correction, including the dependence of the bias signal on outside temperature.}
\label{moon}
\end{centering}
\end{figure*}

\section{Moon effect and CCD non-linearity}

Since early 2016 the Auger FRAM has started taking data also during the presence of the Moon on the sky and it quickly became apparent that such data show a different VAOD distribution from those taken during moonless nights. This effect has been also confirmed by the CTA FRAM (top row of Fig.~\ref{moon}). Moreover, a less prominent effect of similar kind is seen when scans are taken by the Auger FRAM in the direction of the nearby city of Malarg\"{u}e. This points to a (spurious) dependence of the results on the sky background -- however attempts to include such an effect in the model have resulted in neither an improvement of the fit nor the disappearance of the Moon effect. 

It turns out that the likely culprit is the non-linearity of small signals of the CCD cameras we use (Moravian Instruments G4-16000). While this value is perfectly within the expected performance standards of less than 1\% of the dynamical range (it is less than 0.5 \% for any camera we measured), this translates to up to 30 \% of the signal itself for small signals. A typical observer using such a camera chooses the exposure so that the signal from the object of interest falls somewhere in the middle of the dynamic range, and thus experiences a non-linearity of less than 1 \% of the signal, whereas in FRAM operation, where a large number of stars need to be captured in a reasonable time frame, the signal of many individual stars reaches only a small fraction of the whole dynamical range, in particular at large airmasses where extinction is significant. In this range, the behavior of the cameras was not previously characterized in detail and our laboratory measurements \cite{Sergey} have shown that it is complex and varies from camera to camera.

\begin{figure*}
\begin{centering}
\includegraphics[width=\textwidth]{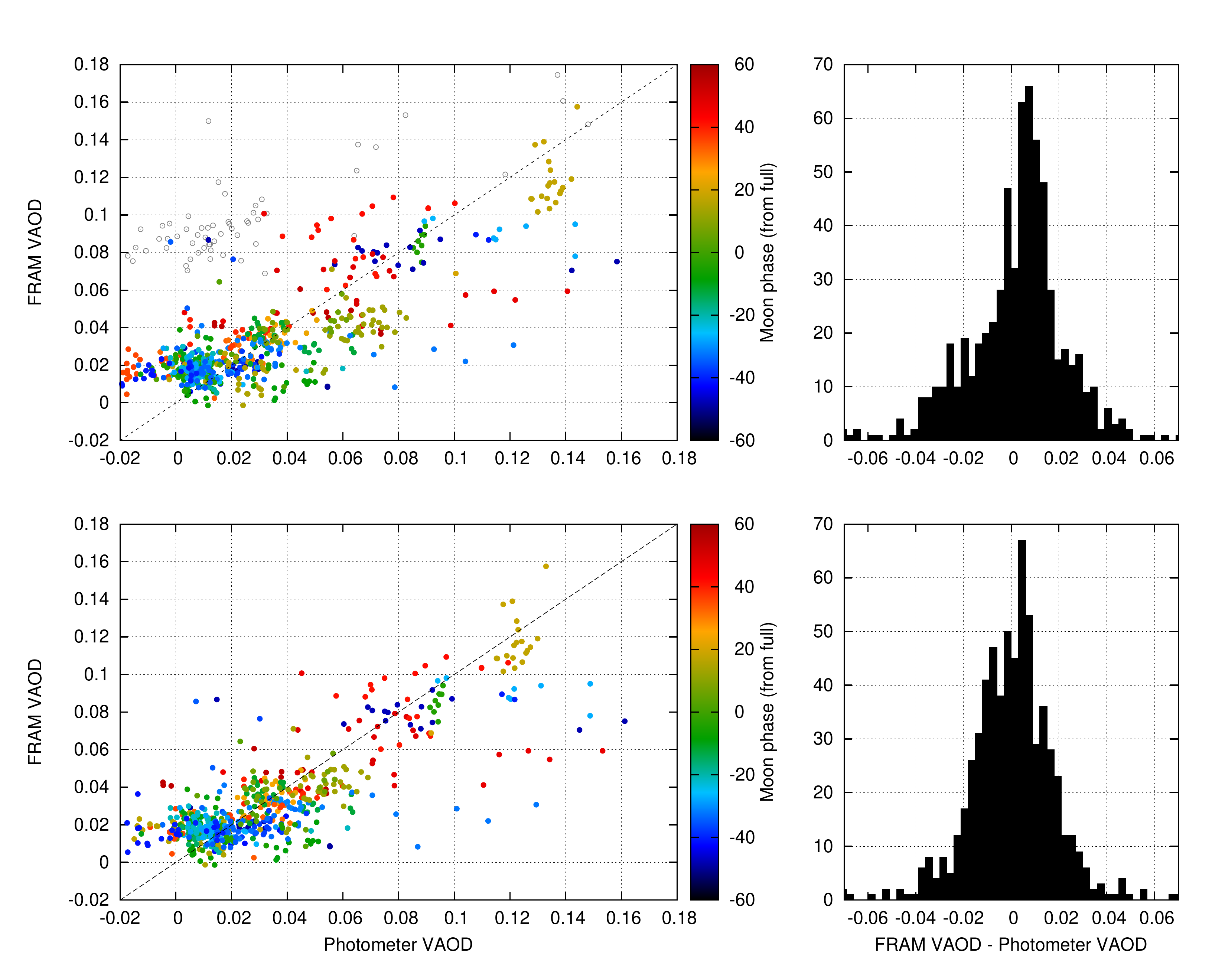}
\caption{Comparison between Sun/Moon Photometer and CTA FRAM VAOD measurements within 15 minutes from each other. Each point represents a FRAM measurement; several more Photometer measurements usually fall within 15 minutes and their values are averaged. The top row shows the comparison and histogram of differences for original Photometer data, the bottom row shows the effect of a fitted correction for Moon phase. The empty gray circles in the upper left plot show the excluded FRAM outliers that are not considered in the histogram or the Moon phase fit.}
\label{foto}
\end{centering}
\end{figure*}

We have thus applied the measured non-linearity corrections (NLC) to all VAOD data. Because this correction is large and rapidly changes for very small signal values, care must be taken to properly subtract the bias signal from each frame, so that only actual light signal is fed to the NLC. This bias signal depends on the temperature of the readout electronics of the camera, which is not actively controlled (unlike the CCD chip itself), but is measured and reported for each image. Using thousands of dark images for each camera, we have created calibration functions which describe the bias signal reasonably well; recently we have implemented the readout of masked regions of the CCD chips of the cameras so that we can measure the bias signal for each frame directly. Using the NLC without the temperature bias correction renders the resulting data unsuitable for VAOD analysis altogether. For both Auger and CTA FRAMs the NLC (with temperature bias correction) removes most of the Moon effect (Fig.~\ref{moon}) even though the Moon-affected data show a slightly larger spread. Note that, somewhat counter-intuitively, the NLC mostly changes the Moonless data, as when Moon is present, the sky background shifts the pixel levels into higher signal values where the non-linearity is smaller.

The $M$ parameter in the model should have in principle compensated for a small non-linearity. In reality, the non-linearity at pixel level translates into a different effect for each individual star depending on the point spread function (PSF) of the star which is influenced by focus, position in field of view and any adjustment errors. Interestingly, for CTA FRAM, the apparent non-linearity determined from the fit ($M-1$) decreases from 2-6 \% without the NLC to <0.5 \% with the NLC, while for the Auger FRAM, it is only modestly affected -- this again shows that correcting for this effect at magnitude level is already too late in the processing chain. It is also important to note that the non-linearity has been discovered independently (when searching for variable objects in the data archive) and it cannot be easily inferred from the VAOD fits; the NLC does not appreciably change the quality of the fit. The RMS of the spread of individual stars around the fitted model is typically around 0.1 mag (which is consistent with the published accuracy of the Tycho2 catalog) and thus an effect of the scale of approximately 0.02 mag  may not be apparent. Thus a naive determination of the VAOD from the fit may lead to a wrong conclusion even though no obvious discrepancy (besides the Moon effect) is observed.

\section {Precision of the measurements}

The statistical error in the determination of the VAOD from an individual scan is between 0.003 and 0.008 for the Auger FRAM and between 0.002 and 0.004 for the CTA FRAM which can, due to its larger field of view, cover a larger span in airmass in a given time window. The precision is thus largely determined by systematics. Several sources are easily quantified: the subtraction of the molecular atmosphere (0.003, shall be improved using the MODTRAN software and GDAS atmospheric profiles \cite{GDAS}), the uncertainty in determining the system parameters from the fit (0.003), the freedom of choice of the cutoffs in brightness and airmass (0.005). Further sources are yet to be quantified: the precision of the measurements of the system spectral response, the quality of the bias correction, possible biases in the stellar catalog, the influence of the automatic rejection of outlying stars, residual cloud contamination, etc. The effect causing the systematic outliers shown in \cite{PJ} also needs to be  identified and either removed or included in the systematics (unless it is physical, which is unlikely, see next section). Any possible significant improvement in the systematics will inevitably mean that the statistical error becomes dominant -- this is however to a large extent likely a feature of the Tycho2 catalog and the attempts to use different catalogs have been met with mixed results so far due to coverage inhomogenity (APASS) or very different spectral bands (GAIA) and thus this might be the ultimate limit of the method until a better all-sky photometric catalog is available.

\section {Sun/Moon photometer comparison}

Comparison of the FRAM VAOD data with other instruments, such as LIDARs, is notoriously difficult because the FRAM measures the integral extinction all the way through the atmosphere. However the Sun/Moon Photometer Cimel CE318-T, installed next to the CTA FRAM in Chile as a part of the same site-chracterisation campaign, does measure the same quantity using the  Sun and the Moon \cite{Jakub} and it is thus ideal for this comparison. Unfortunately, due to communication problems on the remote location, the data from this instrument could be retrieved only after a recent on-site visit and thus the data processing presented here is extremely preliminary. 

Fig.~\ref{foto} shows the comparison between FRAM and Photometer measurements. The Photometer measures VAOD in many narrow-band filters from which we have chosen the 440 nm one which is very close (within 10 nm) to the effective mean wavelength of all FRAM setups. The outliers in the FRAM data shown in \cite{PJ} were removed for this purpose. The mean value for $VAOD_\mathrm{FRAM}-VAOD_\mathrm{Photometer}$ is 0.009 with 68 \% of the data contained within $\pm0.02$, which is highly encouraging since the possible uncertainty of the Moon measurements of the Photometer is quoted as 0.04 (in contrast to <0.01 for Sun measurements). Nevertheless, an obvious dependence on the Moon phase is visible in the data (across several lunar cycles), pointing to a small deficiency in the Moon illumination model. We have fitted two separate fourth-order polynomials to the difference between the FRAM and Photometer data for waxing and gibbous Moon (as there is a large step in this difference around full Moon). Applying this correction reduces scatter of the data and 68~\% of the differences is now contained within $\pm0.016$ (with the mean value being 0 by construction, see bottom row of Fig.~\ref{foto}).

The Sun/Moon Photometer was also taken for three months in 2017 to Argentina, where it performed a dedicated measurement campaign next to the Auger FRAM. Unfortunately, the weather in this period was less than optimal and the amount of useful data taken is small. The data also show an apparent dependence on lunar phase, but because for many phases only one data point was obtained, it is not possible to do a similar fit. We cannot use the correction obtained from the CTA FRAM as the Photometer is calibrated for each site individually using the respective data.

\section*{Acknowledgments}
The authors are grateful for the support by the grants of the Ministry of Education, Youth and Sports of the Czech Republic, projects LTT17006, LM2015046, LTT18004, LM2015038, CZ.02.1.01/0.0/0.0/16\_013/0001402, CZ. 02.1.01/0.0/0.0/16\_013/0001403 and CZ.02.1.01/0.0/0.0 /15\_003/0000437. This research was supported in part by the Polish National Science Centre under grant 2017/26/D/ST9/00449. The work is partially performed according to the Russian Government Program of Competitive Growth of Kazan Federal University.

\end{document}